\documentclass[runningheads]{llncs}
\usepackage[fontsize=9.8]{scrextend}

\usepackage{cite}
\usepackage{amsmath,amssymb,amsfonts}
\usepackage{multirow}
\usepackage{algorithmic}
\usepackage{graphicx}
\usepackage{textcomp}
\usepackage{balance}
\usepackage[utf8x]{inputenc}
\usepackage[table,xcdraw]{xcolor}
\usepackage[pdfa]{hyperref}
\usepackage{csquotes}
\usepackage[english]{babel}
\usepackage{algorithmic}
\usepackage{graphicx}
\usepackage{textcomp}
\usepackage{paralist}
\usepackage{slashbox}
\usepackage{url}

\usepackage[inline,shortlabels]{enumitem}

\setlist[itemize]{noitemsep, topsep=0pt}
\setlist[enumerate]{noitemsep, topsep=0pt}
\def\mysize{9pt}
\begin{document}
\title{Variational Autoencoders for Anomaly Detection in Respiratory Sounds}
%
%
\author{Michele Cozzatti \and
  Federico Simonetta\orcidID{0000-0002-5928-9836} \and
  Stavros Ntalampiras\orcidID{0000-0003-3482-9215}}
\authorrunning{M. Cozzatti et al.}
%
\institute{LIM -- Music Informatics Laboratory\\
  Department of Computer Science\\
  University of Milano\\
  \email{michele.cozzatti@studenti.unimi.it, stavros.ntalampiras@unimi.it}  \\ \url{https://www.lim.di.unimi.it/}}
\maketitle              

\begin{abstract}

  This paper proposes a weakly-supervised machine learning-based approach aiming at a tool to alert patients about possible respiratory diseases.
  Various types of pathologies may affect the respiratory system, potentially
  leading to severe diseases and, in certain cases, death. In general,
  effective prevention practices are considered as major actors towards the
  improvement of the patient's health condition.
  The proposed method strives to realize an easily accessible tool for the
  automatic diagnosis of respiratory diseases. Specifically, the method leverages
  Variational Autoencoder architectures permitting the usage of training
  pipelines of limited complexity and relatively small-sized datasets. Importantly, it offers an accuracy of
  57\%, which is in line with the existing strongly-supervised approaches.


\end{abstract}

\section{Introduction}

  \begin{figure}
    \centering
    \includegraphics[width=0.75\textwidth]{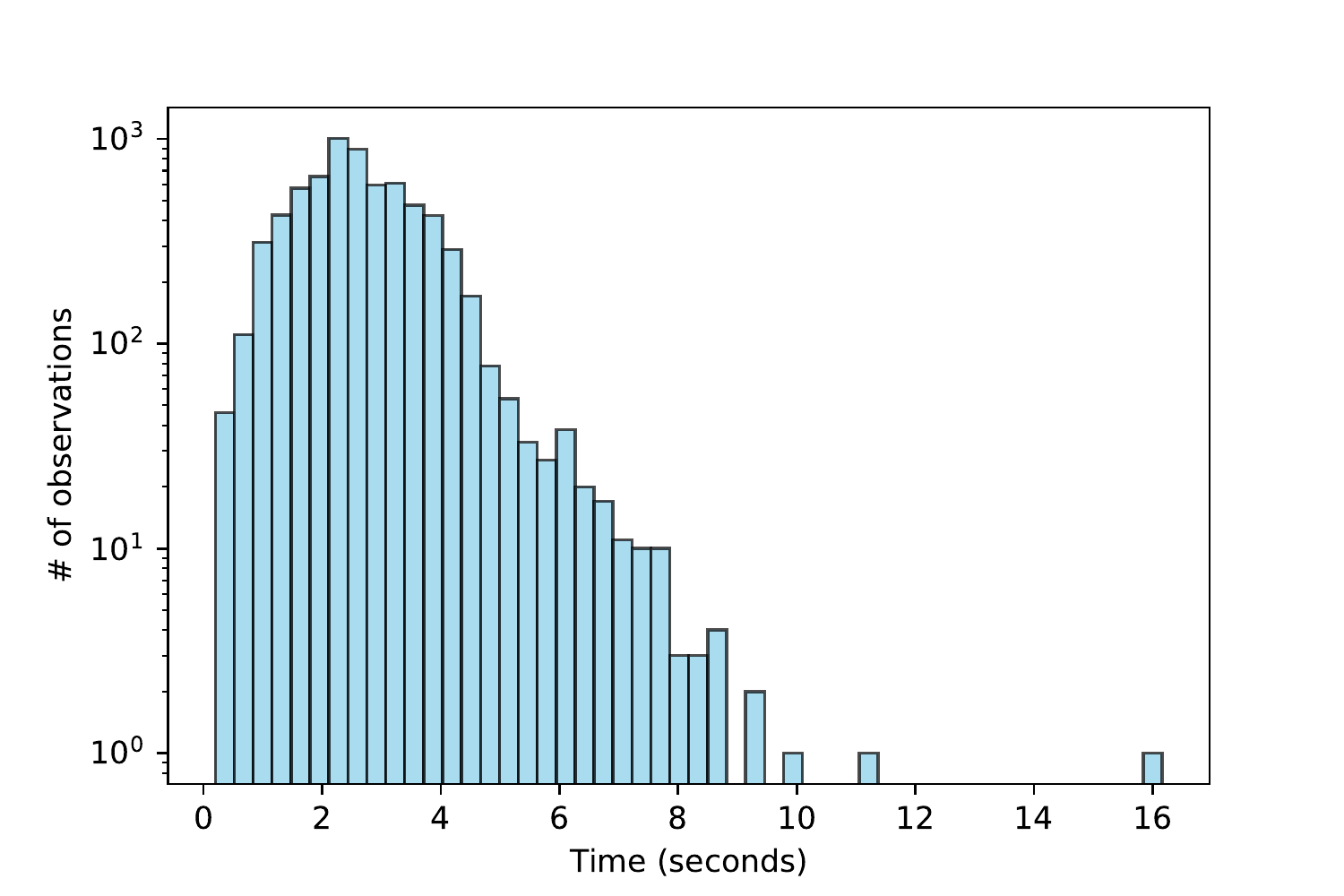}
    \caption{
      Breathing cycle length versus number of observations. The lengths vary
      from a minimum of 0.2 seconds to a maximum of 16 seconds.
    }
    \label{fig:breathing_interval}
  \end{figure}

  The human respiratory system may be affected by pathological conditions which
  typically alter the patterns of the emitted sound events
  \cite{Zak2020}.  Recently, due to the COVID-19 pandemic, such kind of
  diseases became of primary importance for the public health, being a novel
  cause of death for large part of the population \cite{9555564}.
  Interestingly, medical acoustics, i.e. the scientific field using audio signal
  processing and pattern recognition methods for health diagnosis
  \cite{Beach2014}, can play a fundamental role in the development of
  user-friendly tools to prevent and limit the spread of respiratory diseases
  \cite{Ntalampiras2019}.  The ability to create automatic methods to detect
  anomalies is useful for both patients and physicians. If the first ones can
  take advantage from automatic diagnosis methods, physicians can save time and
  minimize potential errors in the performed diagnoses, improving the treatment
  path for the patients. Importantly, such solutions may also reduce the overall
  pressure on public health systems.

  Respiratory/breathing cycle is the result of the combined operation of the
  diaphragm and rib muscles enabling inhalation and exhalation, i.e. breathing in
  and out. Breathing cycles are evaluated by physicians with the aid of a
  stethoscope to spot abnormalities that can represent the onset of a disease.
  Unfortunately, stethoscopes can suffer from external noises, sounds emitted by a variety of organs, etc.

  A respiratory cycle can be classified as normal or abnormal. Abnormal breathing
  cycles can mainly contain two types of abnormal sounds: \textit{crackles}
  and \textit{wheezes}. \textit{Crackles} are mostly found in the
  inhalation phase, although in some cases, they appear in the exhalation phase~\cite{bohadana2014fundamentals}. They are characterized by a short duration and they
  are explosive. \textit{Wheezes} are longer than crackles, around 80-100
  milliseconds, and the frequency can be below 100 Hz. They are usually between
  100 Hz and 1000 Hz and rarely exceed 1000 Hz~\cite{bohadana2014fundamentals}.

  The literature includes various works focusing on the creation of automatic systems that
  can detect abnormalities in respiratory cycles. Usually, feature extraction is
  performed to capture significant properties of the signals (audio features) in
  order to train machine learning models, such as Hidden Markov
  Model~\cite{jakovljevic2017hidden}, Support Vector Machine~\cite{serbes2017automated}
  or boosted decision tree~\cite{chambres2018automatic}. Recently, deep
  learning has been introduced in the audio signal processing community, bringing
  advances to medical acoustics as well. In this case, typical features are
  matrix-like representations of the signal in a frequency-time space. Usually,
  FFT-based representations are extracted and used as an input for training deep
  learning networks, such as MLP~\cite{do2021classification},
  CNN~\cite{perna2018convolutional,demir2019convolutional,do2021classification}, RNN~\cite{perna2019deep}, and other neural
  architectures~\cite{pham2022ensemble,pham2020robust}. In some works, multiple
  type of features are used with exceptional results~\cite{do2021classification,
  tariq2022featurebased}.

  Most of the existing works dealing with the classification of respiratory
  cycles assume availability of data representing every potential class, thus modeling the specific task as a multi-class classification problem. In this paper,
  instead, we propose an anomaly detection approach. The underlying idea is that
  abnormalities in respiratory cycles are difficult to record and can vary
  between different patients, due to different ages and/or sexes. To account for
  such variability, a sophisticated modeling would be required, which is a hard
  requirement given the little amount of available data. Therefore, we decided to
  opt for an anomaly detection task that requires fewer data for training.
  Interestingly, Variational Autoencoders have been successfully employed for
  anomaly/change/novelty detection tasks, such as bird species
  \cite{Ntalampiras2021bird}, ultrasounds \cite{9552041}, time series \cite{Matias2021RobustAD}, computer networks
  \cite{9527980}, etc.

  The main contribution of this article is to investigate how Variational
  Autoencoders (VAE) can be effectively employed for detecting anomalies existing
  in respiratory sounds. The proposed method achieves state-of-art results, while
  only requiring recordings of  normal cycles for training, and as such improving
  considerably the model's generalization abilities. To distinguish the
  abnormalities, a threshold must be set, which can be optimally computed from a
  set of abnormalities. The following sections describe the
  \begin{inparaenum}[a)]
    \item employed dataset of respiratory sounds, \item audio signal preprocessing,
    \item VAE's architecture and training procedure, \item experimental set-up and results, as well as \item our conclusions.
  \end{inparaenum}

\section{The Dataset of Respiratory Sounds}
  \label{sec:dataset}

  We used the Respiratory Sound database provided by the 2017 International
  Conference on Biomedical and Health Informatics
  (ICBHI)~\cite{rocha2019open}. The database consists of 5.5 hours of audio:
  6898 breathing cycles from 920 audio recordings of 126 patients. Audio
  recordings last from 10 to 90 seconds and the audios include different types of
  noises typical of real-world conditions. Moreover, the audio samples were
  acquired using various equipments, sampling rates and detection methods. Each
  respiratory cycle is categorized into four classes: \textit{crackles},
  \textit{wheezes}, \textit{crackles and wheezes}, and \textit{normal}.

  The creators of the challenge predefined train and test subsets of data which
  permits the reliable comparison between different approaches. Such division is
  performed at the level of recordings -- i.e.\ the 920 records from which the
  respiratory cycles are extracted. In this way, a patient’s observations can be
  either in the train set or in the test set, but not in both of them. As such,
  any patient-specific bias is eliminated. The direct consequence is that using
  the predefined division, the generalization abilities of the models are better
  assessed.

  The predefined division is shown in the Table~\ref{tab:ICBHI_60-40}. Overall,
  the train set accounts for the 60\% of the whole dataset, while the test part
  consists of the remaining 40\%. In addition to the predefined division, several
  works in literature used random splits defined at the breathing cycle levels.
  To compare the proposed method to existing works, we tested our method in this
  setting as well. Namely, we used 80\% of the cycles for training and the
  remaining 20\% for testing -- see Table~\ref{tab:ICBHI_80-20}.

  \begin{table}
    \centering
    \footnotesize
    \caption{ICBHI Dataset 60\%/40\% splitting (statistics at the cycle level). Note that in
      the train set abnormal cycles are excluded.}
    \begin{tabular}{|c|c|c|c|c|c|} \cline{2-6}\multicolumn{1}{c|}{} &
               \textbf{Crackles}                                & \textbf{Wheezes} &
               \textbf{Both}                                    & \textbf{Normal}  &
               \textbf{Total}                                                                      \\ \hline \textit{Training} & 962                                                   & 401 & 281 & 1669
                                                                & \textbf{3313}                    \\ \hline \textit{Validation} & 253 & 100          &
               89                                               & 394              & \textbf{829}

               \\ \hline \textit{Testing} & 649  & 385                              &
               143                                              & 1579             & \textbf{2756} \\ \hline
               \textbf{Total}                                   & \textbf{1864}     &
               \textbf{886}                                     & \textbf{506}     &
               \textbf{3642}                                    & \textbf{6898}                    \\
               \hline
    \end{tabular}
    \label{tab:ICBHI_60-40}
  \end{table}

  \begin{table}
    \centering
    \footnotesize
    \caption{ICBHI Dataset 80\%/20\% splitting. Note that in the train set, abnormal
      classes are excluded.}
    \begin{tabular}{|c|c|c|c|c|c|} \cline{2-6}\multicolumn{1}{c|}{} &
               \textbf{Crackles}                                & \textbf{Wheezes} &
               \textbf{Both}                                    & \textbf{Normal}  &
               \textbf{Total}                                                                      \\ \hline \textit{Training} & 1192                                                   & 572 & 325 & 2325
                                                                & \textbf{4414}                    \\ \hline \textit{Validation} & 290 & 144           &
               93                                               & 577              & \textbf{1104}

               \\ \hline \textit{Testing} & 382 & 170                              &
               88                                               & 740              & \textbf{1380} \\ \hline
               \textbf{Total}                                   & \textbf{1864}     &
               \textbf{886}                                     & \textbf{506}     &
               \textbf{3642}                                    & \textbf{6898}                    \\
               \hline
    \end{tabular}
    \label{tab:ICBHI_80-20}
  \end{table}

  Finally, since the proposed method performs anomaly detection, we grouped all
  the anomalies in one class ($anomalies = \{crackles, wheezes, both\}$) and the normal observations
  in another class ($normal = \{normal\}$). Moreover, for the sake of evaluating
  the generalization abilities of the proposed model, we used a validation set
  randomly drawn from the train set. Since the proposed method is weakly
  supervised,  in the training set we disregarded the abnormal respiratory cycles
  and only employed the normal ones. We still used the abnormal cycles in the
  validation and test sets following both of the above-mentioned divisions.

\section{Preprocessing of Respiratory Sounds}

  All recordings were resampled at 4 KHz since, according to the
  literature~\cite{jakovljevic2017hidden, serbes2017automated}, most of the relevant information is below 2
  KHz. As such, according to the Nyquist Theorem, 4 KHz is the minimum-sample
  rate that allows to reconstruct the important information content.
  Subsequently, we  extracted each breathing cycle from each audio file using
  the annotations available in the dataset; in the example shown in
  Table~\ref{tab:annotationsample}, we extracted 9 audio excerpts, disregarding the
  remaining non-useful portions of audio. The available audio excerpts varied in
  duration, with the average being 2.7 seconds. Fig.~\ref{fig:breathing_interval}
  shows the distribution of the obtained audio duration.

  \begin{table}[b]
    \centering
    \footnotesize
    \caption{Example of breathing cycles recorded in one audio file. 0 means
      ``absence'', 1 means ``presence''.}
    \begin{tabular}{|c|c|c|c|c|}
      \hline
      \textbf{ID}       & \textbf{Start}   & \textbf{End} &
      \textbf{Crackles} & \textbf{Wheezes}
      \\ \hline
      1                 & 2.3806           & 5.3323       & 1 & 0 \\ \hline 2 &
      5.3323            & 8.2548           & 0            & 0     \\ \hline 3 &
      8.2548            & 11.081           & 1            & 1     \\ \hline 4 &
      11.081            & 14.226           & 1            & 1     \\ \hline 5 &
      14.226            & 17.415           & 1            & 1     \\ \hline 6 &
      17.415            & 20.565           & 1            & 1     \\ \hline 7 &
      20.565            & 23.681           & 0            & 1     \\ \hline 8 &
      23.681            & 26.874           & 0            & 1     \\ \hline 9 &
      26.874            & 30               & 0            & 1     \\
      \hline
    \end{tabular}
    \label{tab:annotationsample}
  \end{table}

  To ease the processing of the audio excerpts, and similarly to previous works,
  we used a fixed-length for each excerpt \cite{9412226}. Specifically, respiratory cycles
  lasting less than 3 seconds were wrap-padded while those that lasted more 3
  seconds were truncated. To represent the audio characteristics of each excerpt and their evolution in
  time, we extracted Mel-Frequency Cepstral Coefficients (MFCCs) on 
  moving windows. MFCCs are widely used in audio signal processing, including
  speech emotion recognition~\cite{9373397}, acoustic scene
  analysis~\cite{8721379}, music analysis~\cite{simonetta2022context}, and
  medical acoustics~\cite{Ntalampiras2020}.  First, the audio signal was
  converted in a log-amplitude spectrogram using windows lasting 40ms and
  overlapping by 50\%;  Hamming windowing function was applied  before of
  computing the FFT-based log-amplitude spectrum. Then, the Mel filter banks are
  applied to each spectrogram column in order to extract perceptually relevant
  audio characteristics. Finally, the Discrete Cosine Transform (DCT) is used to
  compress the extracted information; the first component, which is proven to be
  highly correlated with the energy of the signal~\cite{zheng2001comparison}, was
  removed to prevent the model of learning information strongly correlated to the recording conditions. As such, 12  MFCCs were considered.

\section{Anomaly Detection Model}

  \begin{figure*}[t]
    \label{fig:schema}
    \includegraphics[width=\textwidth]{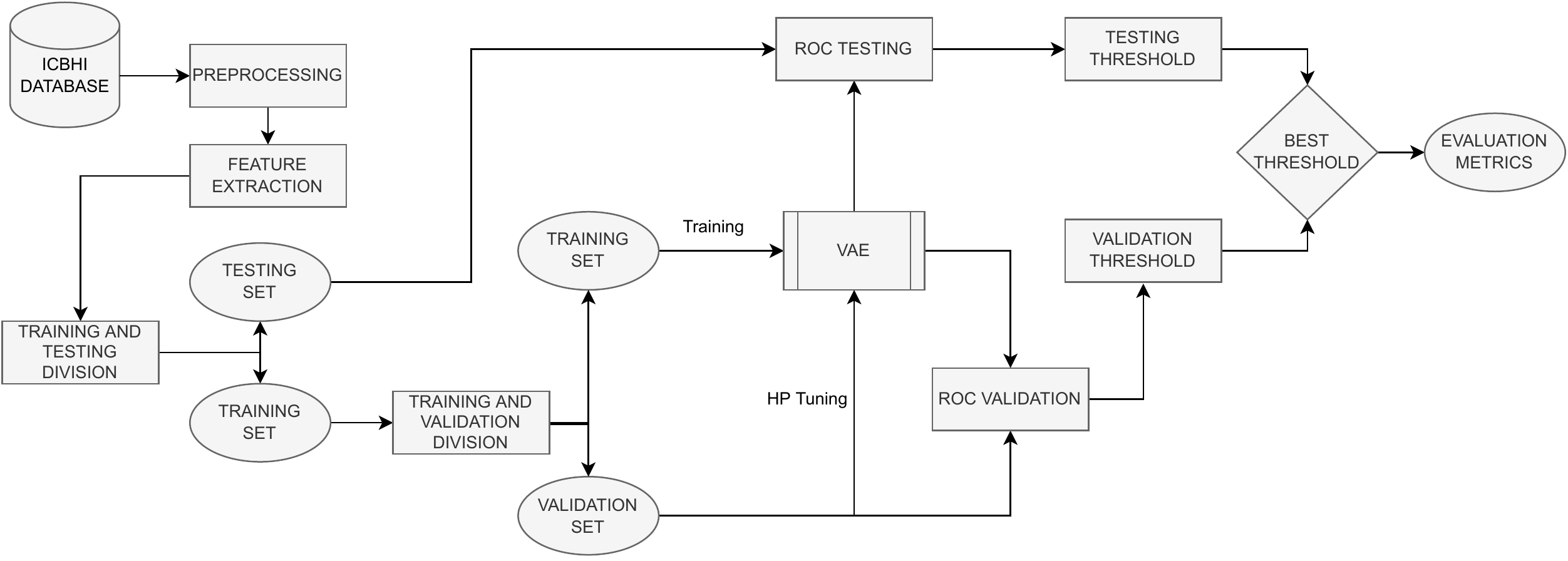}
    \caption{Block diagram of the proposed anomaly detection approach. Starting
      from the database, a pre-processing phase is performed which is followed by
      a feature extraction step. The model is then trained with samples belonging to
      the normal class learning how to reconstruct the input, the optimal
      hyper-parameters are determined and  tested. Finally, the optimal
      thresholds
      for the validation and test sets are determined.}
  \end{figure*}

  Variational Autoencoders (VAE) are neural networks that unify variational
  inference approaches with autoencoders.   Autoencoders are neural networks composed by two parts: the first part, named
  \emph{encoder}, learns a mapping from the input sample
  $x$ to a latent representation $z$; the
  second part, instead, is named \emph{decoder} and learns to map a
  point from $z$ to $x$.

  In variational inference, instead, a distribution is inferred via
  point-estimation methods to approximate Bayesian inference when the number of
  parameters or the size of datasets makes the problem intractable with Monte
  Carlo methods~\cite{blei2017variational}.

\begin{figure}[t]
    \centering
    \includegraphics[width=0.8\textwidth]{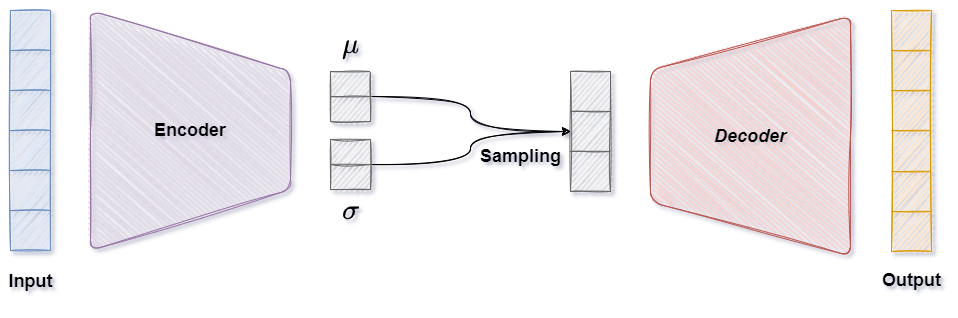}
    \caption{Structure of the proposed VAE consisting of  4 convolutional layers.}
    \label{fig:VAE_structure}
  \end{figure}

  In the case of VAEs, the latent representation $z$ must
  satisfy two important  properties, i.e. it must be
  \begin{inparaenum}[a)]
    \item a continuous distribution, and
    \item easily samplable.
  \end{inparaenum}
  Usually, $z$ is modeled as a Gaussian distribution. In this
  case, for each sample, the encoder predicts a mean and a variance, defining the
  corresponding latent Gaussian distribution. Then, a single point from the
  predicted distribution is sampled and passed to the decoder.
  Figure~\ref{fig:VAE_structure} depicts a VAE with Gaussian latent distribution.

  Compared to basic autoencoders, VAEs allow to draw random samples at inference
  time, making them suitable for generation tasks, such as creativity in music
  applications~\cite{Yamshchikov2020}. Moreover, they allow to mimic Bayesian
  models, which by construction predict distributions. Indeed, when multiple
  samples are drawn from $z$, one can analyze a set of outputs
  that constitute a distribution by themselves, allowing to analyze the epistemic
  certainty measure of the model regarding its own prediction. In other words,
  VAEs can be used to help the final user with an estimation of the certainty of
  the prediction; for instance, a medical expert could decide if the probability
  score predicted by the model should be trusted or not.




  As regards to the loss function, we used the sum of Mean Squared Error between the reconstructed matrix and the input, and the Kullback-Leibler Divergence between the unit Gaussian and the latent distribution.
  We defined each layer as the sequence of a convolutional layer, a
  batch-normalization layer, a ReLU function, and a dropout layer while training.
  The encoder is then built from a sequence of such layers, while the decoder is
  composed by the corresponding layers with transposed convolutions instead of
  simple convolutions. The latent means and standard deviations are computed with
  pure convolutional layers. Figure~\ref{fig:schema} illustrates the whole anomaly detection pipeline.

  \subsection{VAE Training}

    \begin{table}
      \centering
      \footnotesize
      \caption{Hyper-parameters and the optimal value found for each of them. BH indicates the best hyperparameters while HS indicates the hyperparameter space that has been considered.}
      \begin{tabular}{|c|c|c|}
        \cline{2-3}\multicolumn{1}{c|}{} & \textbf{BH} & \textbf{HS} \\ \hline
        \textbf{Loss function}        & MSE + $D_{KL}$  &  \{MAE,MAPE,MSE, MSLE\} \\ \hline
        \textbf{Optimizer}            & Adam  &    \{SGD, RMSprop, Adam, Adadelta\}         \\ \hline
        \textbf{Learning rate}        & $1 \times 10^{-4}$ &  \{$1 \times 10^{-5}$,$1 \times 10^{-4}$,$1 \times 10^{-3}$\}\\ \hline
        \textbf{Epochs}               & 1000      &   -      \\ \hline
        \textbf{Patience}             & 10 epochs  &  \{5,10,15\}      \\ \hline
        \textbf{Batch size}           & 32      &    \{32,64,128\}       \\ \hline
        \textbf{Activation Functions hidden layer} & ReLU  & \{LeakyReLU, ReLU, sigmoid, tanh\}  \\ \hline
        \textbf{Activation Functions output layer} & Linear  & \{linear, sigmoid\}  \\ \hline
        \textbf{Dropout rate}         & 0.3         & \{0.1,0.2,0.3,0.4,0.5\}    \\ \hline
      \end{tabular}
      \label{tab:HP}
    \end{table}

    During training, we first searched for the optimal hyper-parameters using  Bayesian
    Optimization method~\cite{brochu2010tutorial} with a Gaussian Process as
    surrogate model, 2.6 as exploration-exploitation factor $k$,
    and Lower Confidence Bound (LCB) as acquisition function. The hyper-parameters
    and their optimal values are shown in Table~\ref{tab:HP}. We then trained the model using the Adam optimization algorithm with a learning
    rate equal to $1 \times 10^{-4}$.

    Training is performed in a weakly-supervised fashion by using the anomaly
    labels of the validation set only. Specifically, the model is trained to
    reconstruct samples from the normal class, thus when the input is an excerpt
    from the abnormal class, we  expect that the model will not be able to
    efficiently reconstruct the input. Since the network is trained to minimize the
    Mean Squared Error (MSE) between the input and the output, we expect a small
    MSE for normal respiratory cycles and larger MSEs for abnormal ones.
    Consequently, the MSE computed on the training set observations can be used as
    a threshold to spot anomalies.

    Since the dataset is not fully balanced, the threshold is chosen so that it
    maximizes the balanced accuracy on the validation set. Balanced accuracy,
    compared to Matthews Correlation Coefficient \cite{Matthews1975}, allows
    for an easy interpretation, being the average between true-positive-rate (TPR)
    and true-negative-rate (TNR). In our case,    TPR is the rate of correctly
    identified anomalies, while TNR is the rate of correctly identified normal
    observations~\cite{chicco2021matthews,bekkar2013evaluation}.

\section{Experimental Set-Up and Results}

  \begin{table}[t]
    \footnotesize
    \caption{Results obtained on the testing set using two different thresholds,
      one computed on the validation and one computed on the testing set.}
    \label{tab:thresholds}
    \centering
    \begin{tabular}{cc|ccc|}
      \cline{3-5}
      \multirow{2}{*}{}                                              &                                                                & \multicolumn{3}{c|}{Validation threshold}                                  \\ \cline{3-5}
                                                                     &                                                                & \multicolumn{1}{c|}{TPR}                  & \multicolumn{1}{c|}{TNR} & ACC \\ \hline
      \multicolumn{1}{|c|}{\multirow{2}{*}{60\%/40\% split}}             &
      \begin{tabular}[c]{@{}c@{}}Validation\\ threshold\end{tabular} &
      \multicolumn{1}{c|}{0.33}                                      & \multicolumn{1}{c|}{0.80}                                      & 0.57                                                                       \\ \cline{2-5}
      \multicolumn{1}{|c|}{}                                         &
      \begin{tabular}[c]{@{}c@{}}Test\\ threshold\end{tabular}       &
      \multicolumn{1}{c|}{0.44}                                      & \multicolumn{1}{c|}{0.70}                                      & 0.57                                                                       \\ \hline
      \multicolumn{1}{|c|}{\multirow{2}{*}{80\%/20\% split}}             & \begin{tabular}[c]{@{}c@{}}Validation\\ threshold\end{tabular} &
      \multicolumn{1}{c|}{0.48}                                      & \multicolumn{1}{c|}{0.71}                                      & 0.60                                                                       \\ \cline{2-5}
      \multicolumn{1}{|c|}{}                                         &
      \begin{tabular}[c]{@{}c@{}}Test\\ threshold\end{tabular}       &
      \multicolumn{1}{c|}{0.58}                                      & \multicolumn{1}{c|}{0.61}                                      & 0.60                                                                       \\ \hline
    \end{tabular}

  \end{table}

  In the context of the ICBHI challenge, the multi-class accuracy is computed as the
  average of TPR for each class. Having combined all the anomalous
  classes into one, we calculated TPR for two classes (\emph{anomalies}
  and \emph{normals}) and from there, the ICBHI score, which corresponds
  to the balanced accuracy. Moreover, we		  assessed the performance of
  the model using ROC curves and AUC values which are well-established figures of
  merit in the related literature.

\begin{figure}[t]
    \centering
    \includegraphics[width=\textwidth]{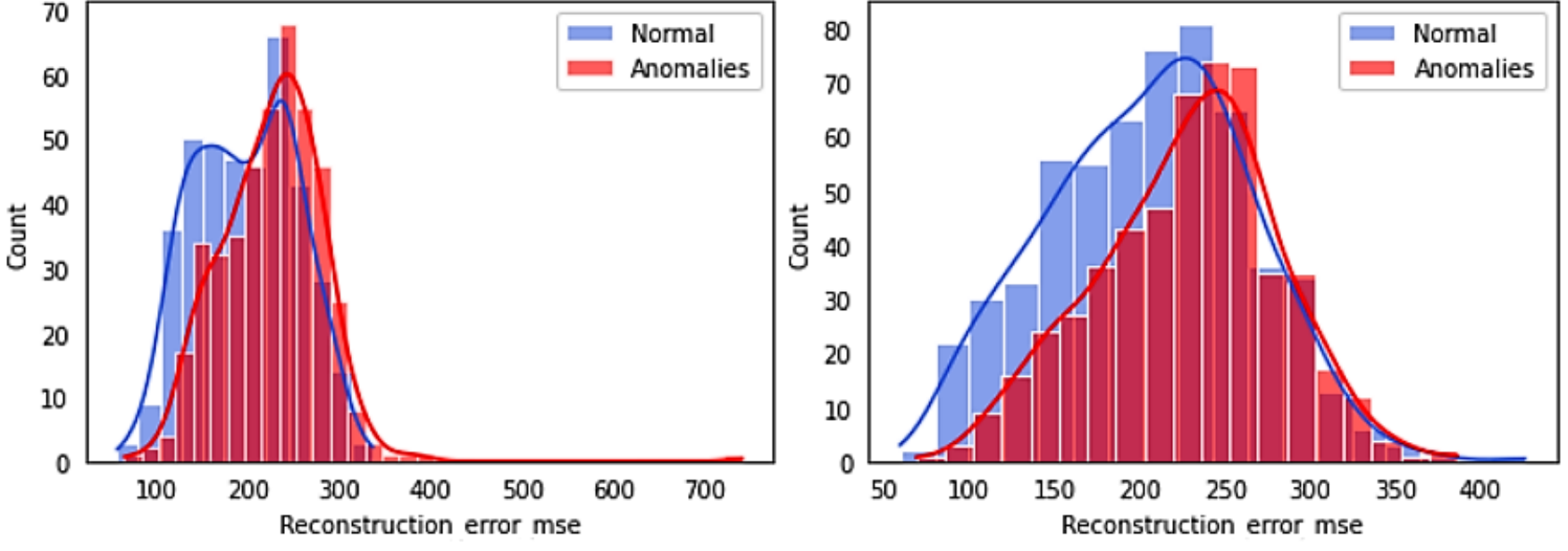}
    \caption{
    \label{fig:Validation}
      Distributions of normal and anomalous samples existing in the validation set. The figure on the
      left is related to the division in 60\%/40\%, while the one on the right in  80\%/20\%.
    }

    \centering
    \includegraphics[width=\textwidth]{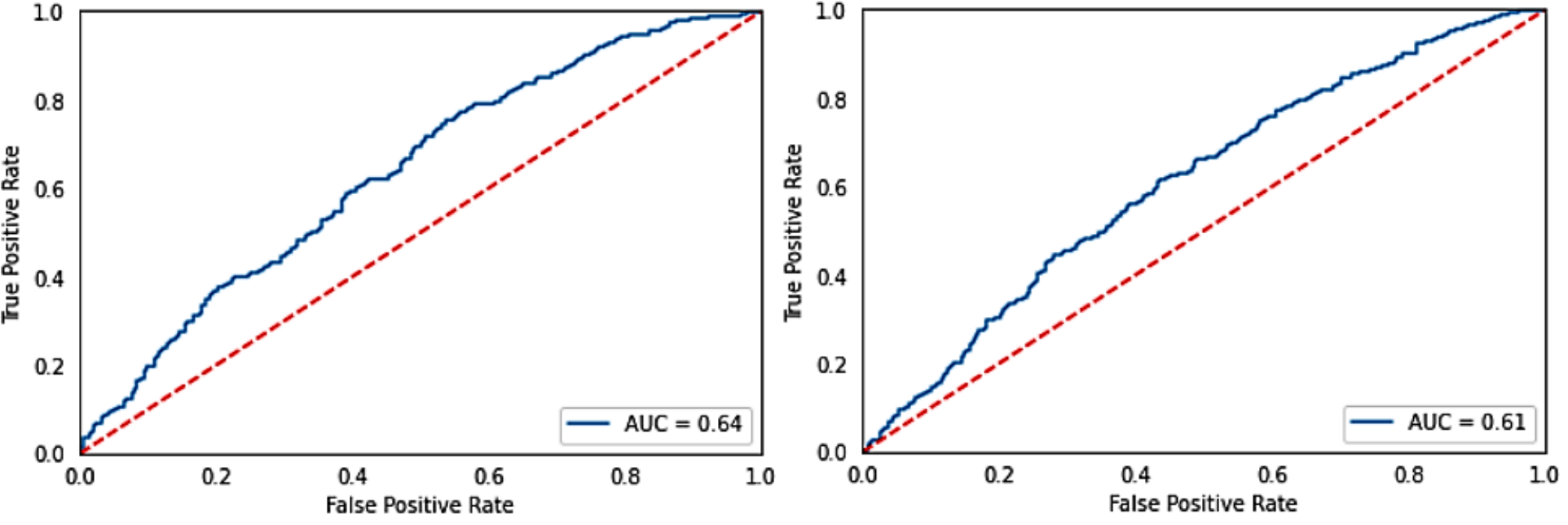}
    \caption{
        \label{fig:ROC_VAL}
      ROC and AUC computed on the validation set. The left image is related to the 60\%/40\% division, while the one on the right to the  80\%/20\% one.
    }
    
  \end{figure}

  We first observed the distribution of the MSE values in the validation set --
  see Fig.~\ref{fig:Validation} -- finding that MSE allowed to partially
  separate excerpts coming from the two classes. We also observed the ROC and AUC while
  training, discovering that an optimal threshold could successfully separate the
  two classes. Fig.~\ref{fig:ROC_VAL} shows the performance of the trained
  model on the validation set.

\begin{figure}[t]
    \centering
    \includegraphics[width=\textwidth]{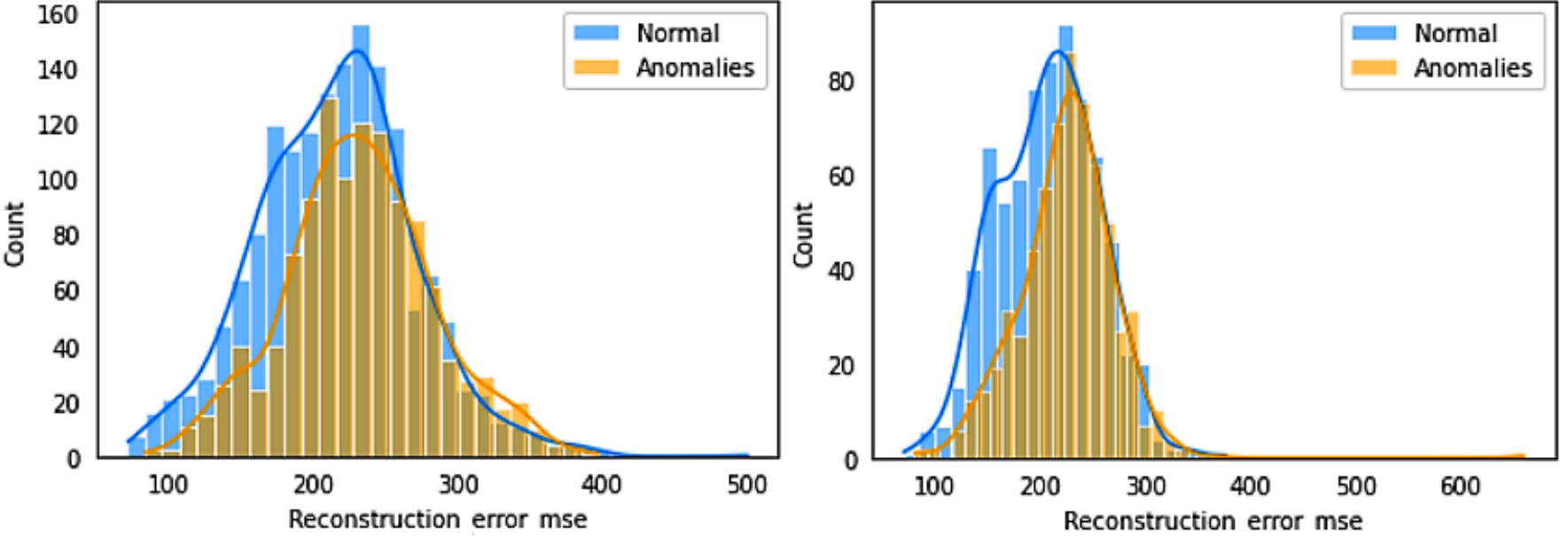}
    \caption{
        \label{fig:Testing}
      Distributions
      of normal and anomalous respiratory sounds as regards to the testing set. The left image is related to
      the division in 60\%/40\%, while the one on the right to the 80\%/20\% one.
    }

    \includegraphics[width=\textwidth]{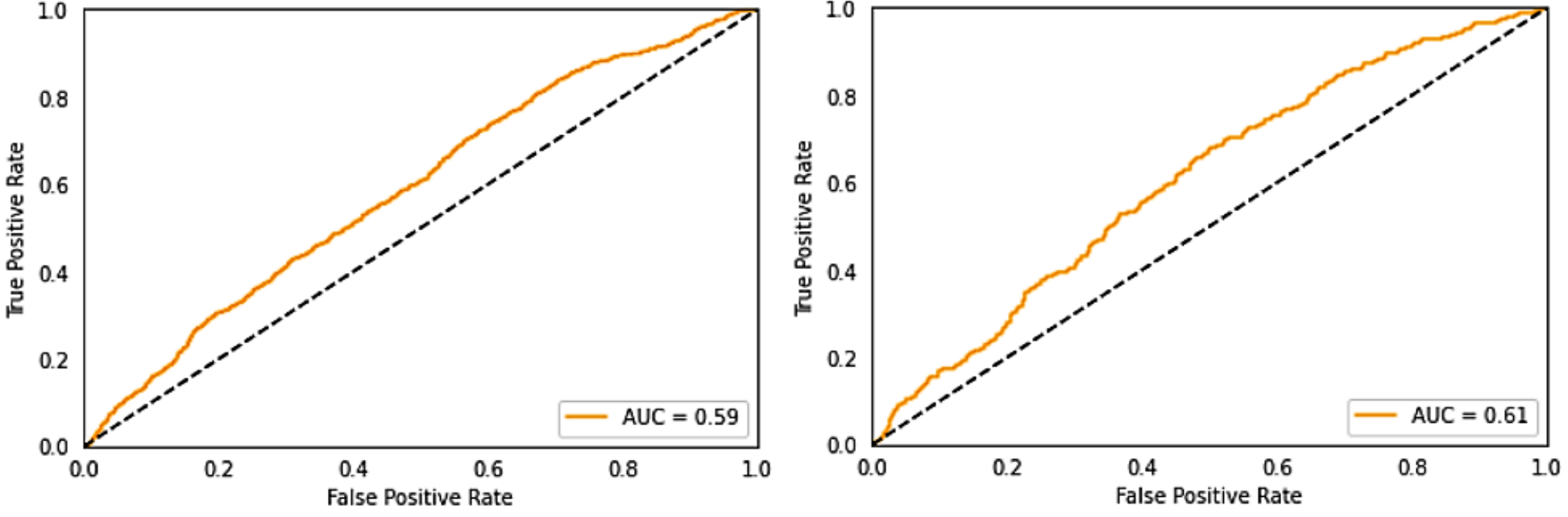}
    \caption{
        \label{fig:ROC_Testing}
      ROC and AUC computed on the testing set. The left image is related to the division in 60\%/40\%, while the one on the right in the 80\%/20\% one.
    }
  \end{figure}

  We performed the same evaluation on the testing set to assess the
  generalization abilities of the model -- see
  Fig.~\ref{fig:Testing}~and~\ref{fig:ROC_Testing}. It is particularly
  interesting comparing the optimal threshold computed on the validation set and
  the corresponding one computed on the testing set. Using the 80\%/20\% division, the
  proposed method identified a slightly smaller threshold than using the
  60\%/40\% division.

  \begin{figure}
    \centering
    \includegraphics[width=\textwidth]{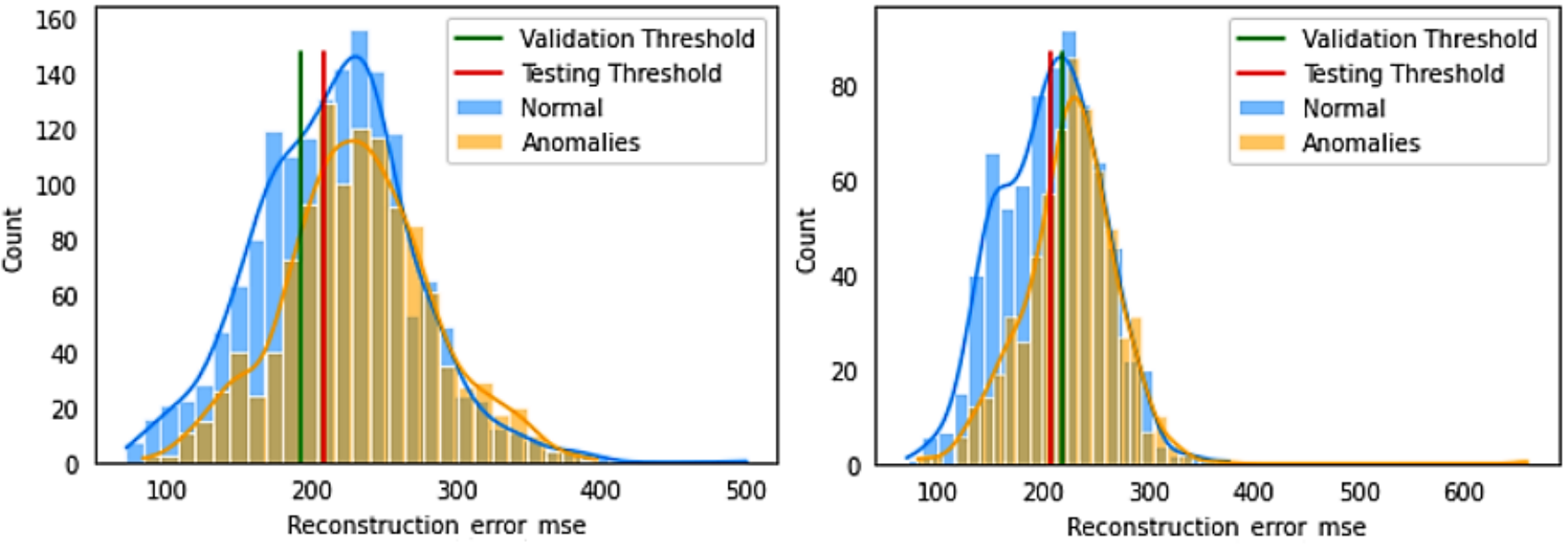}
    \caption{
      MSE thresholds when processing the testing set.
      The left image is related to the division in 60\%/40\%, while the one on the right in the 80\%/20\% one.
    }
    \label{fig:Threshold}
  \end{figure}

  However, by comparing the results obtained using the two thresholds -- see
  Table~\ref{tab:thresholds} -- we observe that the validation threshold is an
  effective approximation of the testing one, proving the generalization
  abilities of the model.



  A fundamental aspect that must be considered when comparing different works is
  the data division into training and testing sets. As explained in
  Section~\ref{sec:dataset}, the creators of the database offer a
  subdivision (60\%/40\%) at the level of audio recordings, so that observations associated with a given patient  can be either in train or test set.
  The division into 80\%/20\% used by some authors is instead extracted randomly
  at the level of the excerpts in which each audio recording is segmented
  (respiratory cycles). Unfortunately, such a division may be biased due to
  patient dependency. Using such a division, the performance measure increases
  essentially for two reasons: 
  \begin{itemize}
      \item there is a greater amount of data in the   training phase, and
      \item observations of the same patient can be both in training
  and in testing.
  \end{itemize}
  Moreover, the usage of random seeds can generate dataset
  division that are particularly successful, thus hiding a pre-bias -- i.e.\
  searching good random splits to boost the final scores.
  
  \begin{table}[t]
    \footnotesize
    \centering
    \caption{
      ICBHI Challenge results on the detection of crackles
      and wheezes (four-class anomaly detection-driven prediction) with the proposed method.
      $TPR$ is the true-positive-rate, $TNR$ is the true-negative-rate and
      $ACC$ is the balanced accuracy (corresponding to the ICBHI score).
    }

    \begin{tabular}{|c|c|c|c|c|c|} \hline \textbf{Method} &
               $\mathbf{TPR}$                         & $\mathbf{TNR}$                      &
               $\mathbf{ACC}$                         & \textbf{Split}                      &
               \textbf{Task}
               \\ \hline HMM~\cite{jakovljevic2017hidden} & 0.38                                & 0.41
                                                      & 0.39                                &
               \multirow{6}{*}{\textit{60/40}}        & \multirow{5}{*}{\textit{4 classes}}

               \\ STFT+Wavelet~\cite{serbes2017automated} & 0.78
                                                      & 0.20                                & 0.49                            &
                                                      &

               \\ Boosted Tree~\cite{chambres2018automatic} & 0.78 & 0.21
                                                      & 0.49                                &
                                                      &
               \\ Ensemble DL~\cite{pham2022ensemble} & 0.86
                                                      & 0.30                                & 0.57
                                                      &                                     &
               \\ \cline{6-6} \textbf{Proposed Method} & 0.33
                                                      & 0.80                                & $ 0.57^*$                       &
                                                      & \textit{2 classes}
               \\ \hline LSTM~\cite{perna2019deep} & 0.85 & 0.62
                                                      & 0.74                                & \multirow{7}{*}{\textit{80/20}} &
                  \multirow{3}{*}{\textit{4 classes}}

               \\ CNN-MoE \& C-RNN~\cite{pham2020robust} & 0.86 & 0.73 & 0.80
                                                      &                                     &

               \\ \cline{6-6} LSTM~\cite{perna2019deep} & -
                                                      & -
                                                      & 0.81                                &
                                                      & \multirow{4}{*}{\textit{2 classes}}
               \\ CNN-MoE \&
               C-RNN~\cite{pham2020robust}            & 0.86                                & 0.85                            & 0.86 &  & \\
               \textbf{Proposed Method}               & 0.58                                & 0.61                            &
               $0.60^*$                               &                                     &
               \\ \hline
    \end{tabular}
    \label{tab:ICBHI_competitors}
  \end{table}

  Overall, the proposed approach offers performance which is in line with the
  state of the art even though it employs respiratory sounds representing only
  healthy conditions. Interestingly, such a line of thought addresses the problem
  in a more realistic way since it is unreasonable to assume availability of
  abnormal respiratory sounds representing the entire gamut of such diseases.

\section{Conclusions}

  In this work we presented a framework modeling the MFCCs  extracted from
  healthy respiratory sounds using a fully convolutional Variational Autoencoder.
  To the best of our knowledge, this is the first time that the detection of
  respiratory diseases is faced from an anomaly detection perspective.
  Interestingly, the proposed model achieved state of the art results in a
  patient-independent experimental protocol even though it is only
  weakly-supervised.

  The small size of the available dataset poses the problem of overfitting and
  model generalization abilities. To this end, we employed Variational Inference,
  which allows the model to estimate its own epistemic confidence, in addition to
  the estimation of the anomaly probability.

  In order to improve the performance of the presented anomaly detection system,
  data augmentation could be a fundamental addition, as it will provide
  additional training samples.  Moreover, different architectures could be
  tested including networks able to take into account time dependencies, such as
  attention-based nets and Recurrent Neural Networks. Finally, multiple features
  could be used to improve the reconstruction abilities~\cite{do2021classification,tariq2022featurebased}.

  \balance

  %
  %
  \bibliographystyle{splncs04}
  \bibliography{zotero}
\end{document}